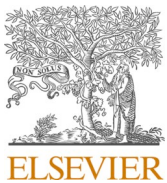
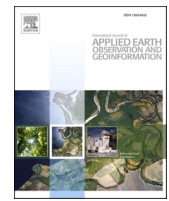

# GlacierNet2: A hybrid Multi-Model learning architecture for alpine glacier mapping

Zhiyuan Xie [a], Umesh K. Haritashya [b,*], Vijayan K. Asari [a], Michael P. Bishop [c], Jeffrey S. Kargel [d], Theus H. Aspiras [a]

[a] *Department of Electrical and Computer Engineering, University of Dayton, Dayton, OH 45469, USA*
[b] *Department of Geology and Environmental Geosciences, University of Dayton, Dayton, OH 45469, USA*
[c] *Department of Geography, Texas A&M University, College Station, TX 77843, USA*
[d] *Planetary Science Institute, Tucson, AZ 85719, USA*

A B S T R A C T

In recent decades, climate change has significantly affected glacier dynamics, resulting in mass loss and an increased risk of glacier-related hazards including supraglacial and proglacial lake development, as well as catastrophic outburst flooding. Rapidly changing conditions dictate the need for continuous and detailed observations and analysis of climate-glacier dynamics. Thematic and quantitative information regarding glacier geometry is fundamental for understanding climate forcing and the sensitivity of glaciers to climate change, however, accurately mapping debris-cover glaciers (DCGs) is notoriously difficult based upon the use of spectral information and conventional machine-learning techniques. The objective of this research is to improve upon an earlier proposed deep-learning-based approach, GlacierNet, which was developed to exploit a convolutional neural-network segmentation model to accurately outline regional DCG ablation zones. Specifically, we developed an enhanced GlacierNet2 architecture that incorporates multiple models, automatic post-processing, and basin-level hydrological flow techniques to improve the mapping of DCGs such that it includes both the ablation and accumulation zones. Experimental evaluations demonstrate that GlacierNet2 improves the estimation of the ablation zone and allows a high level of intersection over union (IOU: 0.8839) score, which is higher than the GlacierNet (IOU: 0.8599). The proposed architecture provides complete glacier (both accumulation and ablation zone) outlines at regional scales, with an overall IOU score of 0.8619. This is a crucial first step in automating complete glacier mapping that can be used for accurate glacier modeling or mass-balance analysis.

## 1. Introduction

The effects of anthropogenic climate change have received considerable attention in recent decades (Anderegg and Goldsmith, 2014; Immerzeel et al., 2010; Johns-Putra, 2016). In particular, rising global temperatures (Houghton, 2005) have resulted in glacier mass loss and terminus fluctuations (Azam et al., 2018; Bolch et al., 2012; Roe et al., 2017; Shrestha and Aryal, 2011; Vuille et al., 2008; Zemp et al., 2015). Observable glacial changes include glacier surging and terminus advancement and retreat, although recession is the predominant trend. Because mountain glaciers are the source for many river systems (Hannah et al., 2007; Immerzeel et al., 2010; Milner et al., 2009), deglaciation can significantly impact montane and downstream ecosystems, as well as human activities related to water supply, hydropower generation, agriculture, industry, and recreation (Huss et al., 2017). Rapid glacial melting can also form high-mountain glacial lakes posing a risk of catastrophic outburst floods for downstream communities (Haritashya et al., 2018; Kääb et al., 2020; National Research Council, 2012; Sattar et al., 2021). Apart from regional effects, mountain glacier mass loss contributes to sea level rise (Chen et al., 2013; Gardner et al., 2013; Radić and Hock, 2011; Zemp et al., 2019). Consequently, it is imperative that we accurately characterize and monitor alpine glaciers at appropriate spatial and temporal scales.

Currently, many glacial studies are focusing on mass balance, glacier volume, and glacier hydrology, which generally require accurate estimates of glacier boundaries and area as one of the primary inputs (https://nsidc.org/data/highmountainasia/data-summaries). Conventionally, these data have been delineated manually on-screen, which requires subjective interpretation to identify boundary even if expert analysts are involved. This qualitative approach introduces large relative errors (precision), as glacial surface features are topographically complex and varied, and the subjective decisions are considerable when different human operators do the mapping. It is especially difficult to distinguish between debris-covered glaciers (DCG) and surrounding sediment and lateral moraines, which maintain spectral similarity in satellite imagery and in some circumstances can have similar topography. In comparison, clean ice glaciers are easier to distinguish and map because ice and surrounding debris are spectrally different.

Other glacier mapping approaches include pixel-, object-, or statistical-based (Bishop et al., 2001; Bonk, 2002; Loibl et al., 2014; Rastner et al., 2013; Robson et al., 2015; Sahu and Gupta, 2018; Smith et al., 2015), which may depend upon satellite imagery and/or land-surface parameters (LSP; sometimes called geomorphometric parameters) generated from a digital elevation model (DEM). The pixel-based approaches were introduced in the 1980s (Gratton et al., 1990; Howarth and Ommanney, 1986; Vohra, 1980) and mainly relied on the statistic or empirical use of visible and near-infrared data to analyze spectral characteristics and assign pixels to different categories (Dobhal et al., 2013; Kayastha et al., 2000; Pratap et al., 2015). However, they were unable to accurately differentiate the margins of DCGs (Alifu et al., 2015; Bhambri and Bolch, 2009; Bolch and Kamp, 2005). In comparison, the object-based approach that incorporates the spatial information (e. g., DEM, LSP, texture feature, etc.) can address the challenging issues (e. g., the debris and glacier under the shadow) and generates slightly better glacier outlines (Rastner et al., 2013). The accuracy of these approaches, however, depends upon how accurately spectral and LSP characterize and differentiate surface-matter variations and topographic properties, and may require moderate to extensive manual post-processing. The pixel- and object-based approaches are similar, utilizing spectral data and process feature information (Alifu et al., 2015; Shukla et al., 2010a), such as indices (Shukla et al., 2018), geomorphometric parameters (Quincey et al., 2014), and band ratios (Kargel et al., 2010). Many spectral methods make use of subjective thresholding that serves as the basis for identifying classification boundaries (Rastner et al., 2013).

Empirical thresholding can remove some elements of variable subjectivity (hence, in ideal circumstances can make the analysis more uniform and more precise), but the very act of thresholding is a subjective decision and can impact accuracy. Furthermore, thresholding cannot comprehensively cover and consider the varied glacial features, especially boundary zones with high geomorphological and spatial complexity, slightly differing vegetation and rock weathering states, differing supraglacial and moraine rock compositions, slightly differing water contents of snow and ice and rock debris, and differing illumination conditions in different images and in different parts of the same image. Hence, accuracy and aspects of precision can be impacted by an inflexible thresholding approach, even though it does remove some differing arbitrary elements of human subjectivity. Without manual, glacier-by-glacier iteration of the thresholding or of the automatic object boundaries, even glacial lakes can be mapped erroneously as damp, dirty ice, or vice versa.

Object-based methods represent a more sophisticated approach for addressing these issues (Rastner et al., 2013) by accounting for spectral homogeneity and topographic structure and spatial information that cannot be accounted for with traditional spectral analysis. Statistical-based approaches also select parameters to establish classification boundaries, although here, they are determined by systematic investigation and evaluation. Unfortunately, many of these approaches have yet to be fully developed, and still require the empirical understanding of data characteristics, required information and substantial expert intervention to generate information and appropriately utilize pattern-recognition algorithms. With adequate expert oversight and editing, results can be favorable (Chen et al., 2020); however, it is a labor-intensive process, and the labor costs limit the geographic areas and number of time stamps that can be mapped in a time series.

With high-frequency, high-resolution global satellite imaging now being undertaken, the manually-intensive methods are absolutely impossible to apply to best effect if large geographic breath, high spatial resolution data, or high temporal resolution—or all three—are needed. The more highly automated methods, however, need validation. Ideally, the goal is to achieve maximum automation with maximum accuracy and precision for widely varying glacier types and locations. The goal is that automated intelligent systems will map glaciers with as much accuracy as a fully expert human could do manually with full access to multispectral images and topography, without human biases, and with the computational speed that could map the whole world of glaciers at high spatial resolution and high temporal frequency.

Recently, research involving glacier mapping is increasingly being done using machine-learning (ML) algorithms (Huang et al., 2014; Karimi et al., 2012; Khan et al., 2020; Zhang et al., 2019), such as artificial neural networks (ANNs), support vector machines (SVMs), and random forest (RF). Similarly, with respect to conventional approaches, these ML algorithms also utilize parameters to construct classification boundaries for separating data; however, instead of user-defined settings, these parameters are automatically defined through continuous, iterative learning of the training data. In this sense, the methods can be like a human continuously adapting the mapping criteria within certain bounds and expertly delineating features—but with a hyperdimensional brain and without the human inconsistencies and time spent on task.

These methods of setting ML algorithm parameters are more objective, more adaptive, and have fewer errors of omission compared to conventional approaches (Janiesch et al., 2021; Javed et al., 2021; Khan et al., 2020; Xie et al., 2021, 2020; Mitchell, 1997). Although the data used in ML-based approaches are similar to those of the more conventional methods, the former approaches are capable of handling more feature data, whereas the latter require a thorough analysis and intellectual understanding of the data, thus inhibiting the effective utilization of additional data types. ML algorithms have high capacity for data selection, limited only by the computational abilities of the machine used; thus, the self-learning properties of these algorithms allow for the understanding and processing of additional data types, supplementing the information for differentiating the glacier boundary from surrounding features.

In addition to the typical ML algorithms that have been used for glacier mapping, several attempts have been made via deep learning (DL) (Lu et al., 2021; Nijhawan et al., 2019, 2018, Xie et al., 2021, 2020); an evolved form of ML primarily based upon neural networks and architectures with more hidden layers. One of the most representative DL algorithms, the convolutional neural network (CNN), has shown outstanding performance in image processing (Alom et al., 2019). The convolutional process connects all elements within a local windowed region each time, thus reducing the number of trainable parameters. Comparatively, typical neural networks generally connect all input elements to each other, so the number of parameters rapidly increases with image size. Accordingly, ML classifiers also adopt a local connection, rather than global, in glacier mapping to avoid massive parameter complexity; this, again, is similar to how a human operates in adapting the mapping criteria, within certain bounds, to differing regions of glaciers that have different characteristics. One limitation of the local connection strategy, however, is the lack of wider neighboring information. CNN indirectly incorporates further pixels by superimposing a small receptive field of many hidden layers for considering neighboring information. That is equivalent to how a human can remember how glaciers express themselves in a different region. The big differences are in the speed, comprehensiveness, and objectivity of the analysis (Janiesch et al., 2021; Javed et al., 2021; Krizhevsky et al., 2012; Mitchell, 1997; Nathani and Singh, 2021; Sarker, 2021; Yasin and Musho, 2020).

The first attempt of CNN image classification was conducted in the 1990s (LeCun et al., 1998); although, the technique was not widely employed until parallel computing support was obtained from graphics processing units in the 2010s (Krizhevsky et al., 2012). Apart from its initial purpose for image classification (He et al., 2016; Sandler et al., 2018; Simonyan and Zisserman, 2014; Szegedy et al., 2015), CNNs have been extended to multiple other uses, including object detection (Girshick, 2015; Girshick et al., 2015; Ren et al., 2015), object tracking (Nam and Han, 2016), and image segmentation (Alom et al., 2018; Badrinarayanan et al., 2017; Chen et al., 2018; Hossain and Chen, 2019; Jégou et al., 2017; Ronneberger et al., 2015), the latter of which is a pixel-wise classification.

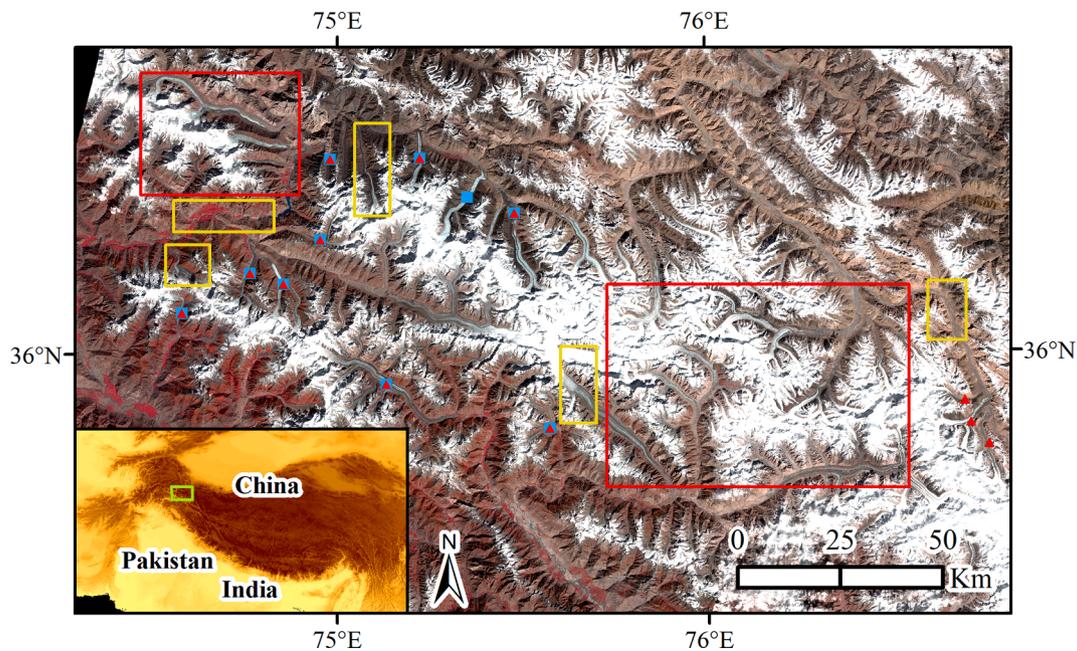

**Fig. 1.** Central Karakoram study area: red and yellow boxes outline the training and validation data regions, respectively. Red triangles mark the glacier for the ablation zone evaluation. Blue squares indicate the location of the glacier for the overall evaluation. Small green box in the inset map identifies the study area in the High Mountain Asia DEM.

We used a CNN segmentation model in developing the DL-based approach, *GlacierNet*, to map DCG ablation zone boundaries in the central Karakoram and Nepal Himalaya (Xie et al., 2020). This approach used limited training data dependent on the size of the total area and can only map the ablation region starting from the snowline to the terminus, often overestimating (i.e., include a few non-glacier pixels) and sometimes underestimating (i.e., excluding a few true glacier pixels) the boundary locations of some terminus. Also, several models have been evaluated and compared to the GlacierNet to find the salient features for improving the glacier mapping algorithm (Xie et al., 2021). GlacierNet and many other automated or semi-automated approach for glacier mapping skips the accumulation zone because of the difficulty in differentiating the accumulation zone snow vs. surrounding snow (Bolch et al., 2007; Lu et al., 2021; Xie et al., 2021, 2020). Therefore, most of the studies that include accumulation zone mapping are either fully or partially done via on-screen manual digitalization (Baraka et al., 2020; Kaushik et al., 2019; Loibl et al., 2014; Mölg et al., 2018; Paul et al., 2009, 2004; Sakai, 2018; Smith et al., 2015; Stokes et al., 2007; Svoboda and Paul, 2009).

The objectives of the present study are 1) to propose a next-generation approach, GlacierNet2, that utilizes multiple models and improves the glacier terminus estimation accuracy, and most importantly, 2) map the snow-covered accumulation zone (SCAZ) and combine it with the ablation zone estimation to produce delineated boundaries for entire DCG.

## 2. Data and methodology

### 2.1. Data

We tested our new approach on glaciers in the central Karakoram in northern Pakistan (Fig. 1), where some glaciers have stable terminus for several decades even though they may be downwasting while others are retreating or surging. This region shows high relief with variable glacier debris cover patterns. Overall it provides a unique test case for wide-ranging glaciers to be tested for automated mapping. This region was also used to test the original GlacierNet algorithm. GlacierNet2 inherited the identical data requirements and processing strategies as GlacierNet as determined by a data layer cross-correlation experiment (Xie et al., 2021, 2020). It comprises all 11 bands of Landsat 8 (https://glovis.usgs.gov), ALOS DEM (30 m), and five layers containing different geomorphometric parameters. Two adjacent Landsat 8 scenes used in this study were acquired on September 24, 2016 (LC81480352016268LGN02) and October 1, 2016 (LC81490352016275LGN01). They contain minimal clouds and fresh snow outside the accumulation zone. These two scenes were mosaicked to cover the study area. It is important to note that September/October time frame is considered the end of the ablation season and most appropriate for mapping glaciers. The geomorphometric parameters were extracted from the ALOS DEM, and includes the unsphericity, profile, and tangential curvatures, as well as slope angle and the slope-azimuth divergence index. Notably, the cross-correlation experiment indicated that none of the 17 data channels were indispensable, but fewer CNN input data channels would lower the accuracy (Xie et al., 2020). It was also shown that all data channels offered complementary advantages, providing unique information for differentiating the DCG ablation zone from the surrounding landscape. Moreover, to avoid information loss and to meet the input requirements of CNN models, all input layers were resampled to 15 m to match the spatial resolution of the Landsat 8 panchromatic band using nearest-neighbor interpolation. Then, the pixel intensities of different input data were normalized to the same order of magnitude based on the numerical value range to prevent the issue of data gradient vanishing.

The DL segmentation technique applied here was a supervised classification method that requires a reference label for training. To compare the GlacierNet2 with the GlacierNet and avoid the performance discrepancies caused by the distinct data, the identical training label data that the GlacierNet has learned are used to train the GlacierNet2. These training labels were converted from the adjusted Global Land Ice Measurements from Space (GLIMS) shapefile (https://www.glims.org). These adjustments include 1) eliminating snow-covered accumulation zones since our previous model focused solely on the DCG ablation zone; 2) re-outline the glacial terminus boundary only if the glacier has advanced or retreated and GLIMS terminus boundary no longer matches on the image being used in this study. In other words, the terminus boundary in the GLIMS database was outlined using a different image, and since then terminus has either retreated or advanced. It should be noted that when GLIMS outlines are overlaid on the images being used in this study, some mismatch was observed on lateral glacier boundaries, possibly due to manual mapping error or glacier downwasting/thickening since mapping was done. However, they were not adjusted and used "as is" because the CNN segmentation model can tolerate a certain degree of error (Xie et al., 2021, 2020; Zlateski et al., 2018). Furthermore, a glacier mapping approach that can generate higher quality estimates through incorporating lower quality boundaries is of significant value for real-world applications and is necessary to address the challenges of deriving error-free boundaries.

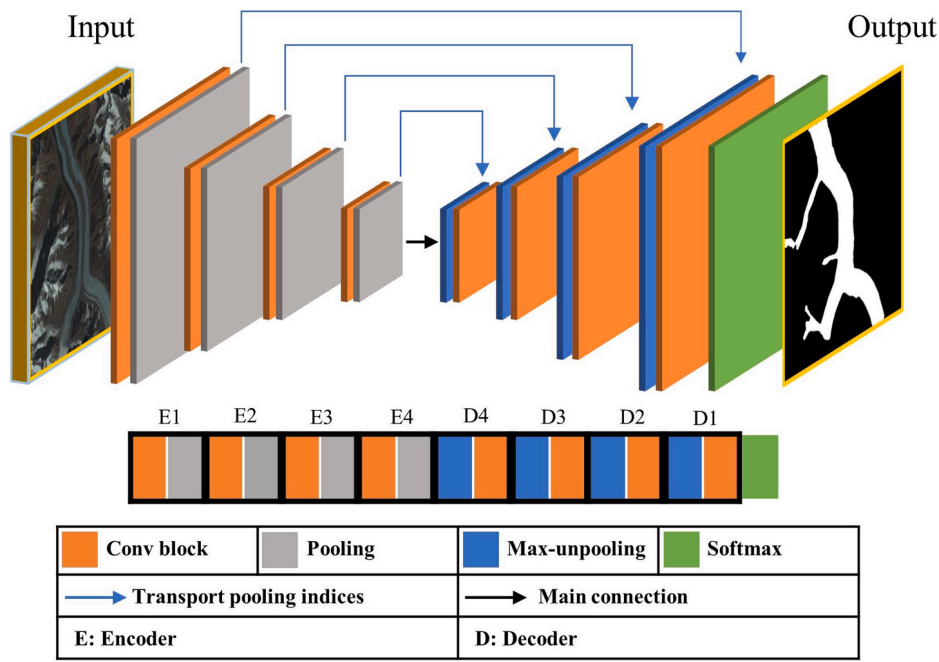

**Fig. 2.** GlacierNet structure.

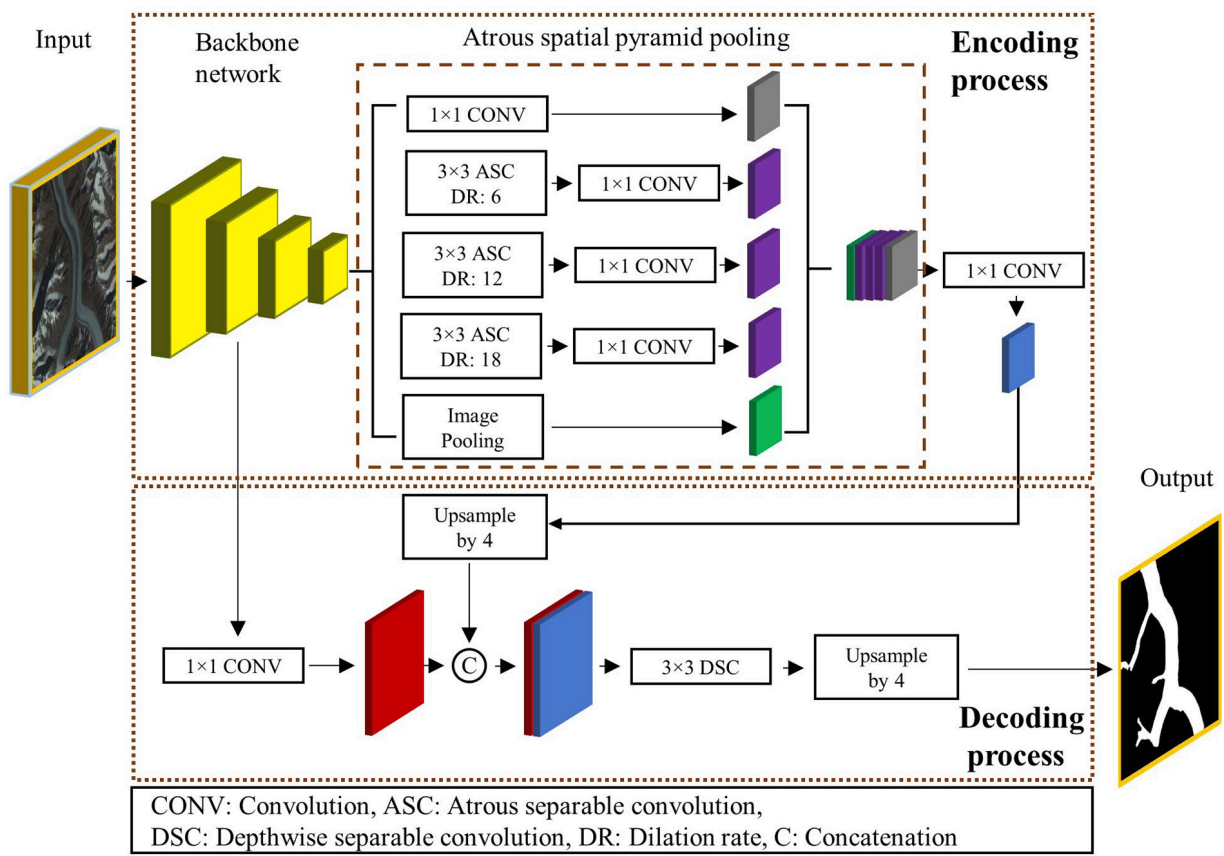

**Fig. 3.** DeepLabV3+ structure.

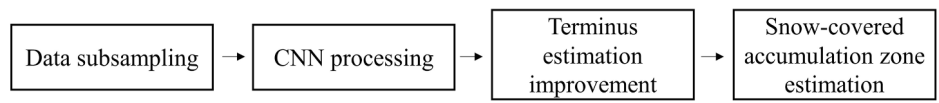

**Fig. 4.** GlacierNet2 brief architecture.

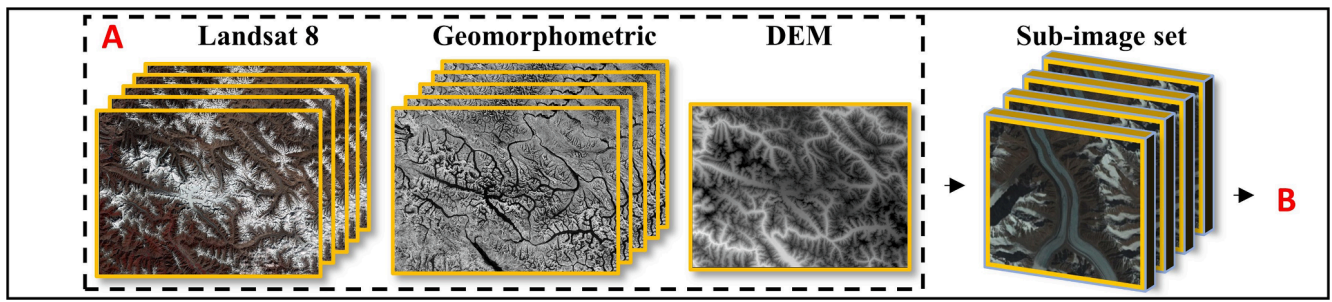

**Fig. 5.** GlacierNet2 Module 1: Data sampling.

The GlacierNet2 performance evaluation also requires the reference boundary. Any error in the reference boundary (including terminus and lateral boundary) can mislead the evaluation results. For the most accurate evaluation, self-generated references are not advised. Since the Glacier Area Mapping for Discharge in Asian Mountains (GAMDAM) dataset has a higher overall quality than GLIMS (primarily due to the same expert delineating all boundaries and errors in all glacier boundaries are similar), the minimally modified GAMDAM boundaries were selected here to evaluate ablation zone mapping (Sakai, 2018). Notably, most parts of these boundaries reasonably match the selected DCGs used for evaluation. It may be because selected central Karakoram glaciers in this study have experienced relatively low retreat compared to the other glaciers in the region. Since GAMDAM data tend to underestimate the SCAZ (Sakai, 2018), the GLIMS-defined SCAZ was merged with the modified GAMDAM ablation zone region to serve as the reference for the overall assessment.

Similar to the GlacierNet approach (Xie et al., 2020), we trained our algorithm using two sub-areas totaling ~15% of the study area, with five smaller partitions designated for validation. Notably, the training sub-areas cover the data from both Landsat 8 scenes used in this study, which allows the network to train and adapt to the reflectance variations that may have been caused by the atmospheric errors. Also, we kept our input simple since the future of this algorithm lies in large-scale glacier mapping, where utilizing hundreds of satellite data without any additional processing would be critical.

### 2.2. Deep learning models

GlacierNet2 modified and utilized two CNN models, including GlacierNet and DeepLabV3+, with the latter being the more complicated of the two. Both networks can produce classification decisions in the form of a binary image that is equal in size to the multichannel input. Furthermore, their input layers are adjusted to fit the 17-channel input rather than the original three channels CNN design for the RGB image.

#### 2.2.1. GlacierNet

GlacierNet (Fig. 2), an axisymmetric structure, was modified from SegNet that incorporated the first 13 convolutional layers from the VGG16 network for its encoding structure (Xie et al., 2021, 2020). The GlacierNet model included equal numbers of encoders and decoders similar to the strategy adopted for SegNet. Encoders reduce the data size and extract outstanding features through the max-pooling layer, whereas decoders employ a reversed max-unpooling to restore data size, and recall the feature back to its original position. Therefore, restoring and recalling processes require pooling indices transported from the corresponding max-pooling layer. In addition to max-pooling or unpooling, both the encoder and decoder have the same major component, known as the convolutional block. This block is constructed by three sequentially connected convolutional layers that contain 32 convolutional kernels with 5 × 5 size. The design strategy is to relax the similarity requirement of the network, thus, successfully marking more positive pixels.

#### 2.2.2. DeepLabV3+

DeepLabV3+ (Fig. 3) is derived from the DeepLab series, representing the most recent revision of DeepLabV3 (Chen et al., 2018, 2017). DeepLabV3+ was constructed by three major modules: the backbone network, atrous spatial pyramid pooling (ASPP), and decoding. The backbone network used is the modified, aligned Xception model (Chen et al., 2018; Chollet, 2017; Qi et al., 2017), which adopts depth-wise separable convolution (DSC) for effectively lowering the computational demand. The DSC, notably different from the conventional convolution, connects each convolutional kernel to a single channel instead of all input data, thus reducing the required computations. Empirically, DSC can achieve near equivalent or slightly improved performances compared to conventional convolution (Chen et al., 2018). Subsequently, the ASPP extracts multi-scale information via five processing chains covering three different methods from the backbone network-produced feature maps. These methods involve the 1 × 1 convolution (point-wise convolution), which is used for decreasing data dimensionality, global average pooling (which averages all elements of each input feature map), and atrous separable convolution (ASC; the fused form of DSC and atrous convolution) with various dilation rates. The atrous convolution (i.e., dilated convolution) inserts holes between the adjacent kernel elements to enlarge the convolutional kernel. Accordingly, the ASC is a special convolution that applies each expanded kernel to the corresponding input channel. Lastly, the decoding module primarily relies upon two bilinear upsampling layers by a factor of four to quickly restore the data to its original size. The bilinear upsampling layer provides the upsampling capacity using the bilinear interpolation. Compared to the transposed convolutional layer, which is also commonly used for upsampling, this layer does not include any trainable parameter and therefore has the advantage of not causing overfitting. Furthermore, effectively utilizing the upsampling layer with a larger upsampling factor (the factor of 4 is larger than the commonly used factor of 2) simplifies and speeds up the decoding process.

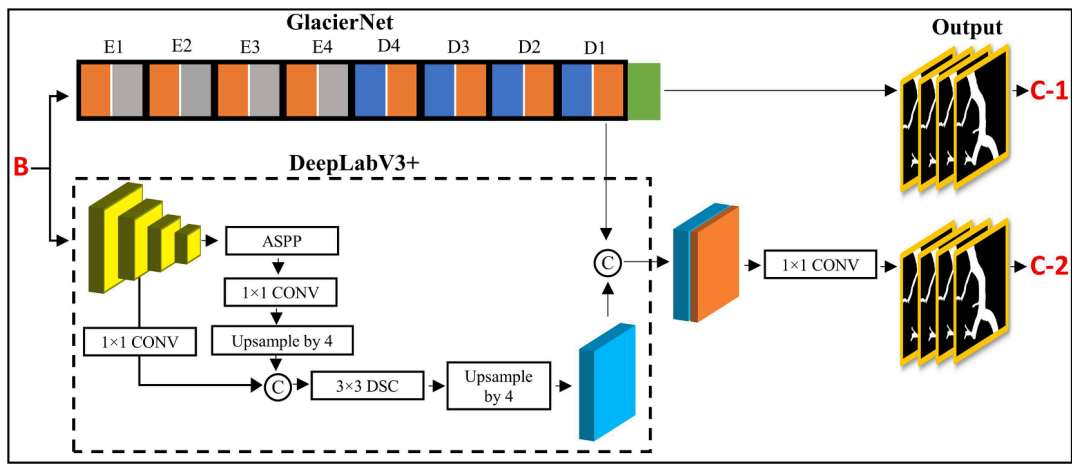

**Fig. 6.** GlacierNet2 Module 2: CNN processing.

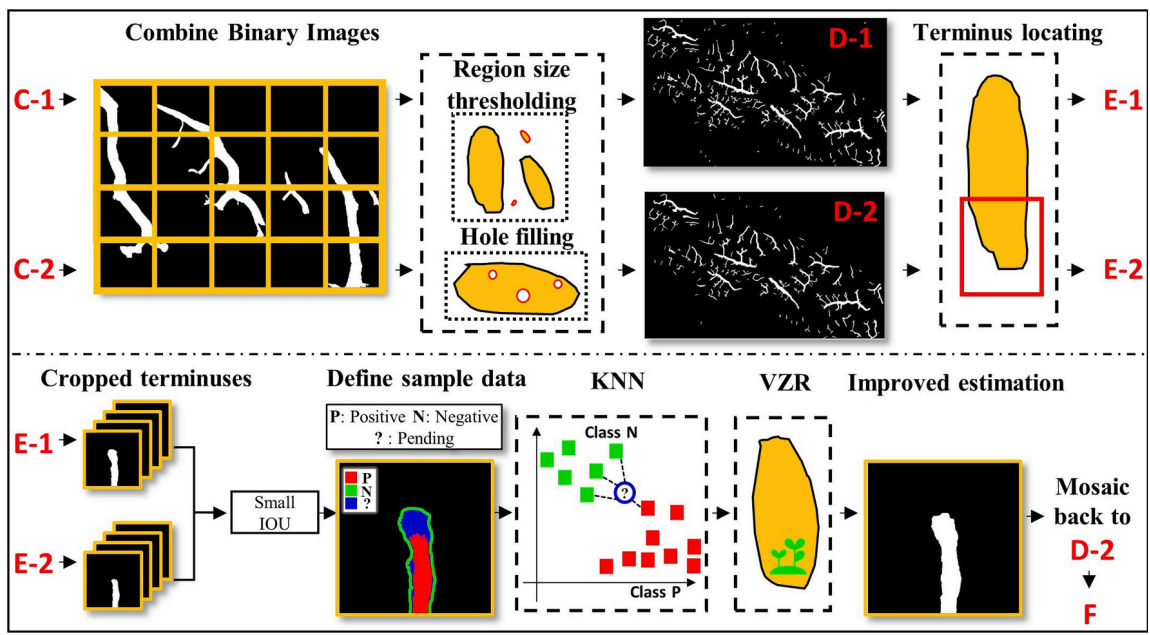

**Fig. 7.** GlacierNet2 Module 3: Terminus estimation improvement.

*2.3. GlacierNet2 architecture*

The overall architecture of the proposed GlacierNet2 consists of four ordered modules for data subsampling, CNN processing, terminus estimation improvement, and snow-covered accumulation zone estimation (Fig. 4). GlacierNet2 generates two types of boundaries, such as those covering the ablation zone and those covering the entire glacier containing both the ablation zone and SCAZ. Below we have discussed each module of GlacierNet2 separately.

*2.3.1. Data subsampling*

The swath of the original multi-channel image is large and computationally intensive if processed as one single image. Therefore, the original image is subsampled using a sliding window with a pixel size of $512 \times 512$ and a shifting stride of 32 (Fig. 5). The window size determines the size of the intermediate feature map being generated by the convolutional layers. The window size utilized in this study is associated with the ASPP of DeepLabV3+, which uses ASCs with large dilation rates; therefore, the efficacy of ASC application on the small size of feature maps is limited.

*2.3.2. CNN processing*

Input data fed into the CNN are processed in parallel through GlacierNet and DeepLabV3+ (Fig. 6). This module exports two intermediate products, including the binary classification GlacierNet, and the final decision incorporates both networks' decisions.

To this end, both networks' feature maps are concatenated and fed into a pointwise convolutional layer for classification. The trainable parameters of the two networks are pre-trained using the training data before being frozen during the training process of the pointwise convolutional layer. In both networks, the final convolutional layer converts the number of feature maps to two for binary classification. GlacierNet exports 32 512 × 512 feature maps from the second to last convolutional layer, while DeepLabV3+ restores the size of the data output from the final convolutional layer through a bilinear upsampling by a factor of four. The second to last convolutional layer outputs 256 feature maps (128 × 128 size). To align data sizes prior to concatenation, these feature maps are expanded via a bilinear upsampling layer. The strategy of using the network fused GlacierNet and DeepLabV3+ is explained in a subsequent section.

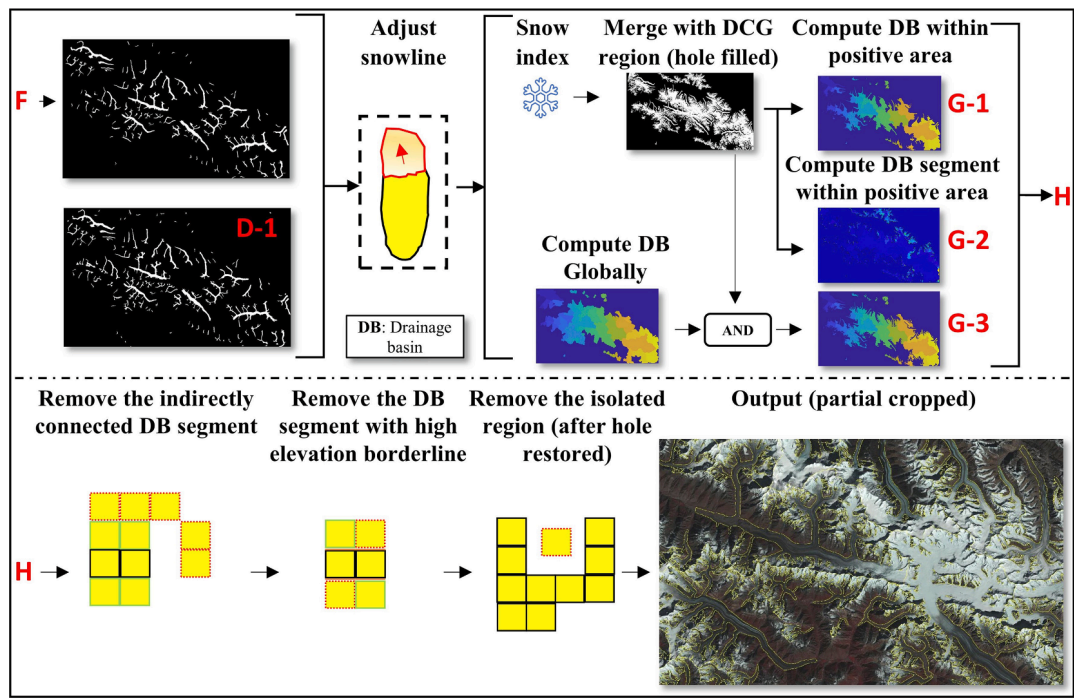

**Fig. 8.** GlacierNet2 Module 4: Snow-covered accumulation zone estimation.

*2.3.3. Terminus estimation improvement*

As the original large-swath satellite imagery was subsampled, two binary image sets from the CNN processing module are combined into large binary images encompassing the entire study area (Fig. 7). Both images are processed separately by region size thresholding to eliminate small misclassified pixel clusters, as well as hole filling to remove small gaps in the ablation zone estimation according to the slope and size thresholding. The previous three steps are inherited from the post-processing of the GlacierNet approach (Xie et al., 2020). The water pixel removal process in the GlacierNet approach is excluded, as the abundance of glacial lakes in the central Karakoram is far less than that in the Nepalese Himalaya.

The additional post-processing steps first detect glacier termini via the DEM using the lowest elevation property of the glacier terminus. On *D-1*, the pixel cluster with the lowest altitude is successively identified for each DCG ablation zone according to the set low altitude percentage (15%, low sensitivity). Based on the pixels identified, the algorithm creates a bounding box to clip the terminus sub-images from *D-1* (GlacierNet estimation) and *D-2* (fused network estimation). It is suggested that the DEM be preprocessed via average filtering to avoid abrupt extreme high or low pixels. Subsequently, the intersection over union (IOU) of two sub-images (Eq. (3)) is calculated. If the IOU is lower than the setting (0.7 here for a low-level of tolerance), the case is defined as one of terminus estimation disagreement. Once a disagreement is identified, the K-nearest neighbor (KNN), a fast non-parametric classification algorithm that requires sample data as a reference (Fix and Hodges, 1989), is used to resolve the classification decision. To effectively define the sample data for KNN, the AND, OR, and XOR results for the two binary sub-images are computed first, where pixels from the AND region are assigned as positive samples, those from the XOR region are pending, and pixels surrounding the OR region are designated as negative. Furthermore, KNN requires input feature data for distance computing and comparison; to this end, output from the first encoder of the GlacierNet is utilized as the input data. Since capturing the sample and feature data required by KNN, a unique KNN classifier can be created for each terminus with estimation disagreement. After applying the KNN to the XOR region (pending pixels), the vegetation zone removing (VZR) process applies the surface reflectance normalized difference vegetation index (NDVI) to remove any vegetated region near the glacial terminus (Eq. (1)):

$$NDVI_{Landsat8} = \frac{Band5_{NIR} - Band4_{Red}}{Band5_{NIR} + Band4_{Red}} \qquad (1)$$

Notably, the KNN results and identified vegetation regions require the image closing operation based on shape and connecting nearby pixels for smoothing edges and removing gaps. Finally, the improved termini are mosaicked back to *D-2*.

*2.3.4. Snow-covered accumulation zone estimation*

The ablation zone from the previous terminus estimation improving module is one of the major inputs for estimating the SCAZ (Fig. 8). The previous CNN processing module incorporate the decision of both DeepLabV3+ and GlacierNet to produce *D-2*. However, disagreement exist between two networks in identifying accurate snowline, causing confusion and producing slightly underestimated *D-2* snowline. In our observation, these disagreements generally occur when some snow pixels are present in the ablation zone. Therefore, the snowline of *D-2* is adjusted to that of *D-1* if the latter has higher elevation values.

Subsequently, the SCAZ estimation employs the TopoToolbox, a powerful MATLAB-based platform with several functions for analyzing terrain, supporting the process of combining spatial and non-spatial numerical analyses (Schwanghart and Kuhn, 2010; Schwanghart and Scherler, 2014). Accordingly, the TopoToolbox program is optimized for processing speed, even when examining highly complex data. It provides a user-friendly and flexible environment for basin geomorphology and hydrology analysis using a DEM. For example, the drainage basin (DB) analysis computes flow direction and accumulation. Our analysis treats each glacier as a DB, which is important for mapping the SCAZ.

**Table 1**
Evaluation indices for debris-covered glacier ablation zone mapping.

| Network | IOU | RC | PC | SP | FM | ACC |
|---|---|---|---|---|---|---|
| GlacierNet (*D-1*) | 0.8599 | 0.8841 | 0.9692 | 0.9919 | 0.9247 | 0.9677 |
| DeepLabV3+ | 0.8623 | 0.8824 | 0.9743 | 0.9933 | 0.9261 | 0.9684 |
| GlacierNet & DeepLabV3+ (*D-2*) | 0.8622 | 0.8780 | **0.9796** | **0.9947** | 0.9260 | 0.9685 |
| GlacierNet2 | **0.8839** | **0.9010** | 0.9790 | 0.9944 | **0.9384** | **0.9735** |

TopoToolbox can estimate the DB based on a DEM with or without a given target region. If a target region is provided, the program can identify all pixels flowing into the basin; whereas if no region has been provided, all DBs defined by the individual hypothetical flow networks are marked. Delineating DBs, however, is not always straightforward as the hypothetical flow network of several individual glaciers almost always converges; therefore, the ablation zones are utilized as a further resource for separating and defining individual glacier basins.

The SCAZ estimation algorithm, apart from the snowline-adjusted *D-2*, requires the DEM and optical data as input (Fig. 8), and the analytical steps were as follows:

1) The surface reflectance snow index (Eq. (2)) is used to identify all snow-covered pixels within the scene:

$$Snow\ index_{Landsat8} = \frac{Band6_{SWIR} - Band3_{Green}}{Band6_{SWIR} + Band3_{Green}} \quad (2)$$

2) Snowy regions are removed if they are not adjacent to and overlapping with the ablation zone.
3) The remaining snowy regions are merged with adjacent or overlapping ablation zone.
4) All gaps in the data are filled, representing the internal, non-snow and non-glacier regions to assist with computational processing.
5) Ablation zones are defined by different code numbers.
6) Pixels are allocated to each target ablation zone, and an intermediate output *G-1* was produced, indicating the DBs of glaciers within the merged region. Each DB subsequently inherits its corresponding ablation zone's code number.
7) The DB for each target ablation zone is computed across the entire scene and also inherits the corresponding ablation zone. Then, the AND operator removes DB pixels located out of the merged region and obtains another intermediate output *G-3*.
8) The intermediate *G-2* is computed, representing the DB segment inside the merged region without a specified target area. Each segment inherits the code number of the corresponding DB from *G-3*. Also, these segments are coded via another independent number series for differentiating.
9) The difference between *G-1* and *G-3* is calculated.
10) The segments of *G-2* corresponding to the differences are marked if it neither neighbored nor overlapped with *G-1* DBs of the same code number.

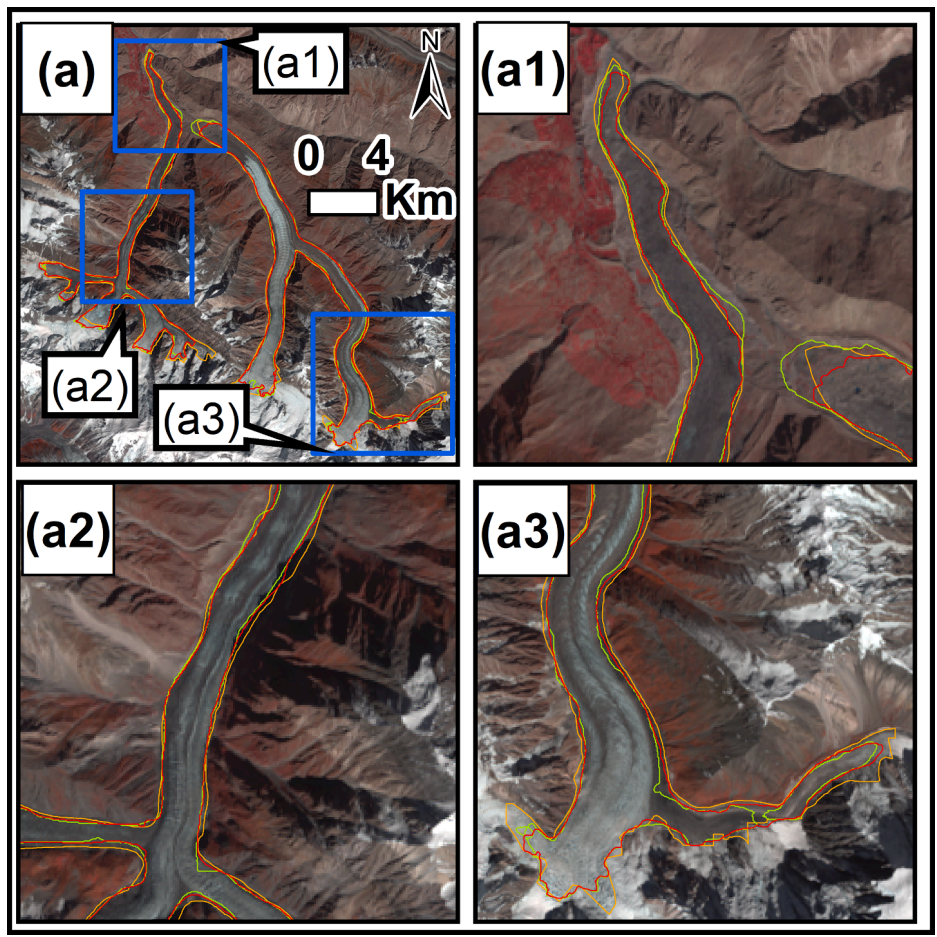

**Fig. 9a.** The ablation zone boundaries of Bualtar and Barpu glacier. The red, green and gold boundaries indicate the GlacierNet2, GlacierNet, and the reference, respectively.

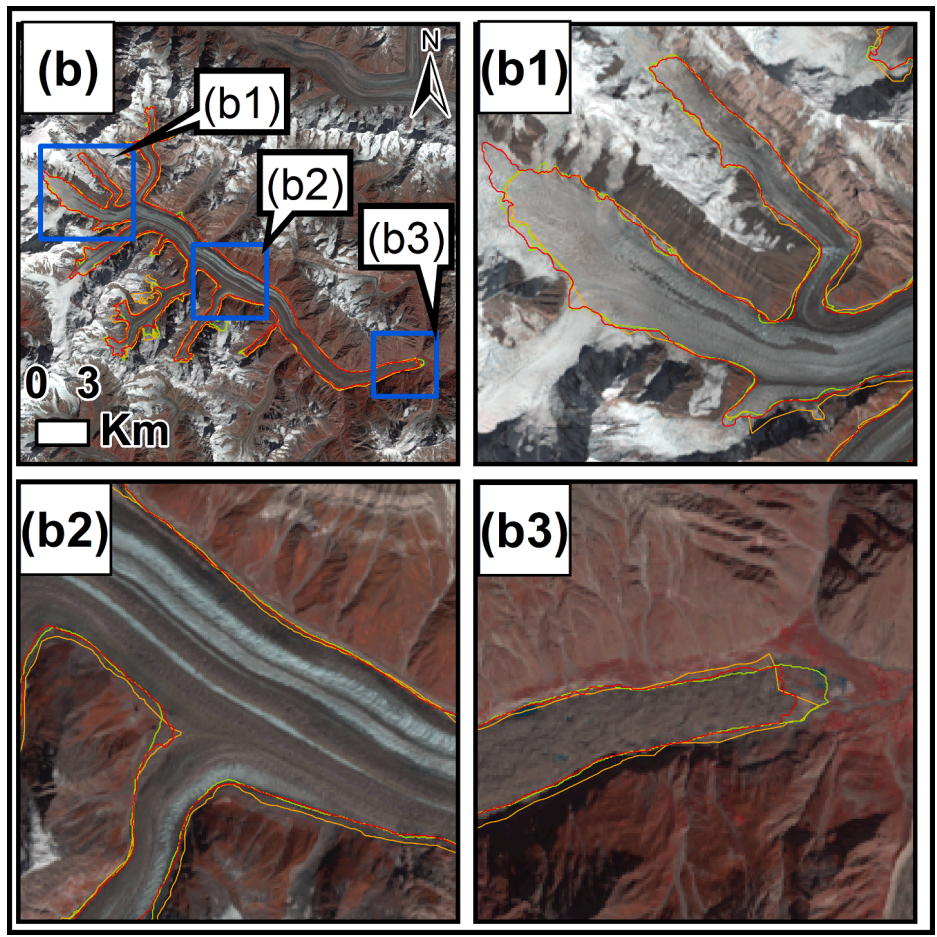

**Fig. 9b.** The ablation zone boundary of Chogo Lungma glacier. The red, green and gold boundaries indicate the GlacierNet2, GlacierNet, and the reference, respectively.

11) The marked DB segments in *G-2* are removed corresponding to the area in *G-3*.
12) The difference between *G-1* and modified *G-3* is recalculated.
13) The segments of *G-2* corresponding to the recalculated differences are identified to locate each segment's connecting borderlines. These connections are part of the segment boundary directly touching the DBs from *G-1*, with the same segment code number. The segments are then selected if the average elevation of its borderline was higher than the mean elevation of the entire segment boundary.
14) The selected segments in *G-2* corresponding to the area in *G-3* are again removed.
15) Morphological close operations are applied to merge the filled gaps from Step 4. Accordingly, all holes (save for the small ones) are restored.
16) Isolated SCAZ regions caused by hole restoration are removed. All isolated regions were independent pixel clusters, unconnected with the corresponding primary glacier region.

Above, *G-1* is slightly underestimated, whereas *G-3* is overestimated, and the total area of *G-2* is equal to that of *G-3*. Indeed, *G-1* accurately estimates most glaciers, although underestimations resulted from prominent moraine ridges on a glacier surface. Accordingly, *G-2* and *G-3* are utilized for refining the estimate further. Also, the merged region is utilized to separate the hypothetical flow network and block the DB segments that indirectly flow into the ablation zone.

### 2.4. Evaluation method

To accurately evaluate DCG ablation zone mapping performance and compare it with other CNN models, we selected the same 12 glaciers distributed throughout the study area as in Xie et al., (2021) (Fig. 1). The selected samples were located outside of the training and validation regions, contained cloud-free ablation zones, as well as varied attributes of size, aspect, surface, and topographical characteristics. For comparison, the assessment scores for GlacierNet, DeepLabV3+, and fused network models based on the GlacierNet approach were calculated as well. The ablation zone evaluation assessed the mapping performance of both the DCG terminus and ablation zone (including lateral boundaries). However, mapping complexities between two areas are different and has been treated separately in this paper. Following the GlacierNet evaluation strategy, a slightly underestimated snowline (SUS) was also introduced for calculating evaluation indices. The SUS is set according to the GlacierNet strategy and is implemented to prevent the effect of snowline delineation error caused by widely distributed snow and mixed pixels. Three glaciers out of 12 used for ablation zone evaluation contains some cloud in the SCAZ region; therefore, for the overall assessment we used 9 out of those 12 glaciers. We, however, also used 1 additional glacier not being used earlier for the ablation zone assessment to test the validity of the network (Fig. 1). The evaluation indices selected for ablation zone and overall assessment included IOU, recall (RC), specificity (SP), precision (PC), F-measure (FM), and accuracy (ACC). They were calculated according to Eqs. (3)–(8), respectively:

$$IOU = \frac{TP}{TP+FP+FN} \text{ or } IOU(Z_1, Z_2) = \frac{R_1 \cap R_2}{R_1 \cup R_2} \qquad (3)$$

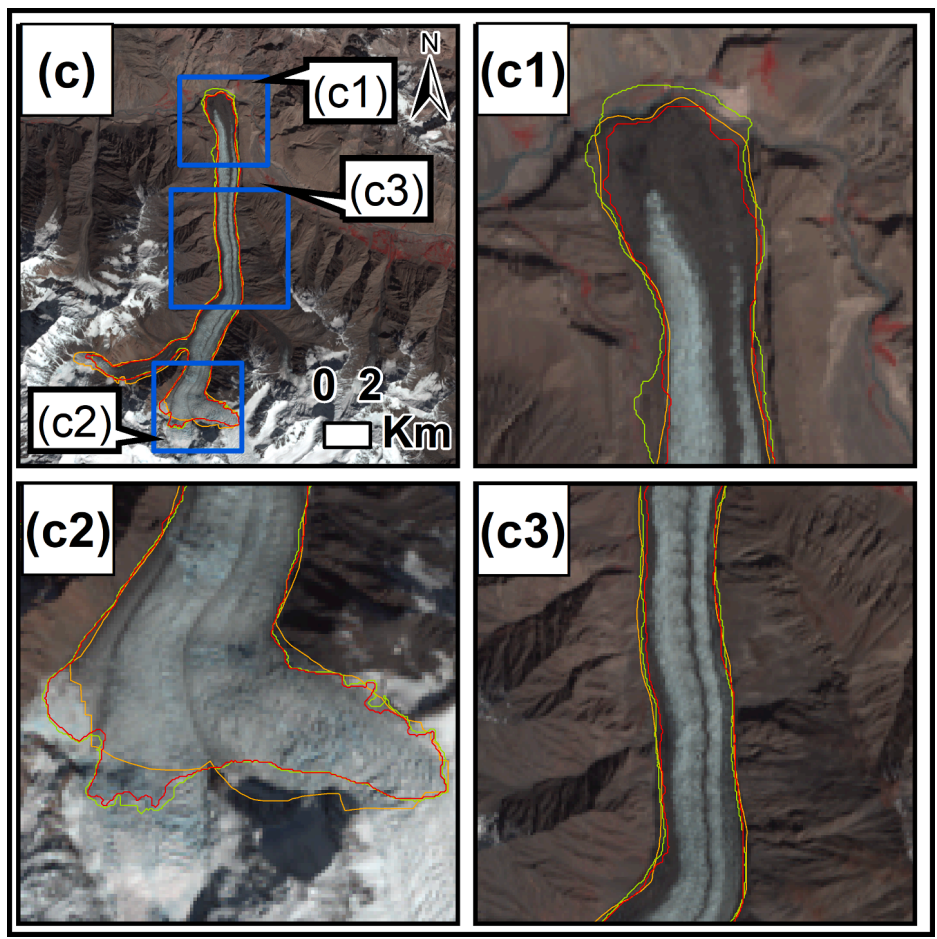

**Fig. 9c.** The ablation zone boundary of Mulungutti glacier. The red, green and gold boundaries indicate the GlacierNet2, GlacierNet, and the reference, respectively.

$$RC = \frac{TP}{TP + FN} \quad (4)$$

$$SP = \frac{TN}{TN + FP} \quad (5)$$

$$PC = \frac{TP}{TP + FP} \quad (6)$$

$$FM = 2 \times \frac{PC \times RC}{PC + RC} \quad (7)$$

$$ACC = \frac{TP + TN}{TP + TN + FN + FP} \quad (8)$$

where TP, TN, FP, and FN represent the numbers of true positive, true negative, false positive, and false negative cases, respectively; while $R_1$ and $R_2$ indicate the two binary pixel regions.

Glacier mapping is an uneven, two-class, pixel-based classification, as the background pixels occupied a significant proportion of the scene, potentially increasing the numbers of true negative cases. The majority of such cases, however, are insignificant due to their distance from the selected glacier. Including these distant background pixels in specificity and accuracy indices significantly narrows the differences in numerical values compared to other approaches. Consequently, the bounding polygons were applied to index calculations for excluding such distant background pixels.

## 3. Results and discussion

In this study, the first assessment examined the performance of four ablation zone estimations, whereas the second assessment examined the results of complete glacier estimation by GlacierNet2. The ablation zone mapping assessment compared the GlacierNet2 approach to the GlacierNet, DeepLabV3+, and fused network (GlacierNet & DeepLabV3+ ). As reported earlier, when GlacierNet was compared with five different CNN models using the same evaluation criteria used here, the GlacierNet and DeepLabV3+ showed higher accuracy (Xie et al., 2021). Therefore, these two approaches were used in this study for comparison. Furthermore, the results of GlacierNet and fused networks are the intermediate outputs of *D-1* and *D-2* in GlacierNet2, respectively.

In ablation zone mapping assessment, GlacierNet2 was the overall top performer, with the highest IOU, recall, F-measure, and accuracy scores (Table 1). The remainder of the IOU scoring rank was as follows: DeepLabV3+ > *D-2* > GlacierNet. Here, *D-2* maintained a similar IOU score as DeepLabV3+ and also produced the highest precision score, which indicates that *D-2* generated the lowest proportion of false-negatives. In addition to the evaluation indices, several DCG ablation zone samples were provided to help elucidate any partial improvements between the two generations of the GlacierNet series (Figs. 9a–9d). GlacierNet2 performed strongly in generating glacier boundaries in the ablation zone, particularly for terminus estimation, representing the most notable improvement in performance over GlacierNet. Additionally, GlacierNet2 performance for lateral boundaries in the ablation zone was superior and generated accurate results except at some snowline regions as discussed in section 2.4 (Fig. 9b1).

For complete glacier mapping, GlacierNet2 delineated the snow-covered area and ridgelines with the highest level of accuracy, resulting in an overall IOU, recall, precision, specificity, F-measure, and accuracy of 0.8619, 0.9299, 0.9218, 0.9603, 0.9258, and 0.9501, respectively (Figs. 10a–10d). Higher recall and lower-precision scores reflect the tendency of GlacierNet2 to overestimate the SCAZ. Some overestimations are caused due to differences in how reference boundaries (GLIMS) have identified SCAZ. For example, some considered only snow on top of the ice to be SCAZ, while others considered the entire side valley walls as part of the SCAZ. As discussed in section 2.3.4, the GlacierNet2 SCAZ mainly accounts for the snowy region that is part of the accumulation zone and directly flows into the ablation zone.

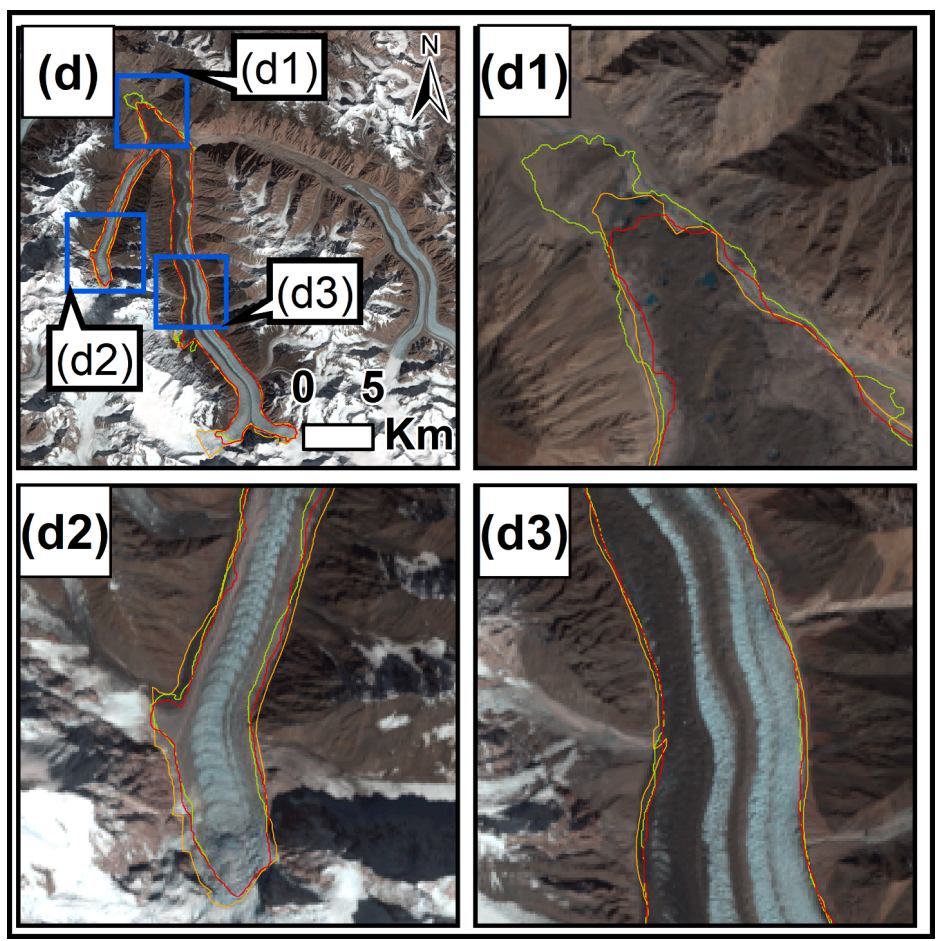

**Fig. 9d.** The ablation zone boundary of Khurdopin glacier. The red, green and gold boundaries indicate the GlacierNet2, GlacierNet, and the reference, respectively.

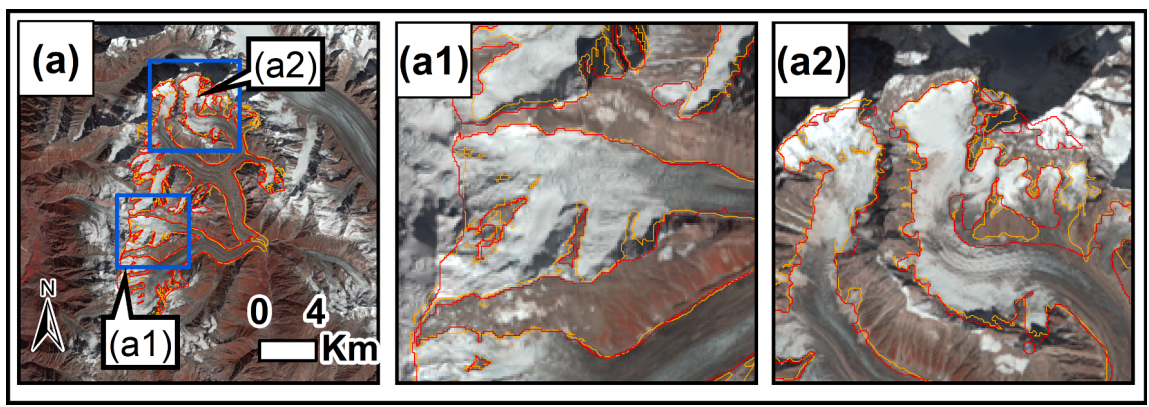

**Fig. 10a.** The boundary covering both ablation zone and SCAZ of the Sosbun glacier. The final output was processed by region size thresholding to remove residuals of small (i.e., few pixels), non-snow areas. The red and gold boundaries indicate the GlacierNet2 and reference, respectively.

Some differences between GlacierNet2 predictions and reference boundaries are caused by the nearby non-snow areas merging into the main glacier and small non-snow region removing processes, as discussed in step 15. The pixels connecting the nearby non–snow areas are the disagreement regions compared to the reference (Fig. 10a2, Fig. 10c3). Furthermore, the small non-snow regions, such as those marked by reference outlines in Fig. 10b1, Fig. 10d1, d2 and those marked by GlacierNet2 in Fig. 14b, were removed and became the differences between the GlacierNet2 and reference.

Furthermore, we have measured the time consumption of four modules including 2.5 h for data sampling and CNN processing, 125 s for terminus estimation improvement, and 229 s for snow-covered accumulation zone estimation, based on the Core i9-10900F processor and GeForce RTX 2080 Ti graphics processing unit. There is a large space for improvement in processing speed which is also part of our future works.

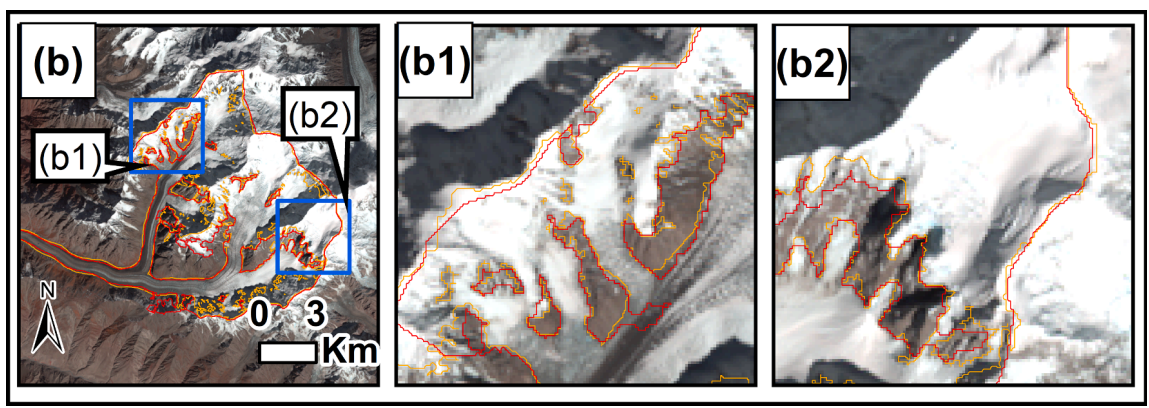

**Fig. 10b.** The boundary covering both ablation zone and SCAZ of Gharesa (Trivor) glacier. The final output was processed by region size thresholding to remove residuals of small (i.e., few pixels), non-snow areas. The red and gold boundaries indicate the GlacierNet2 and reference, respectively.

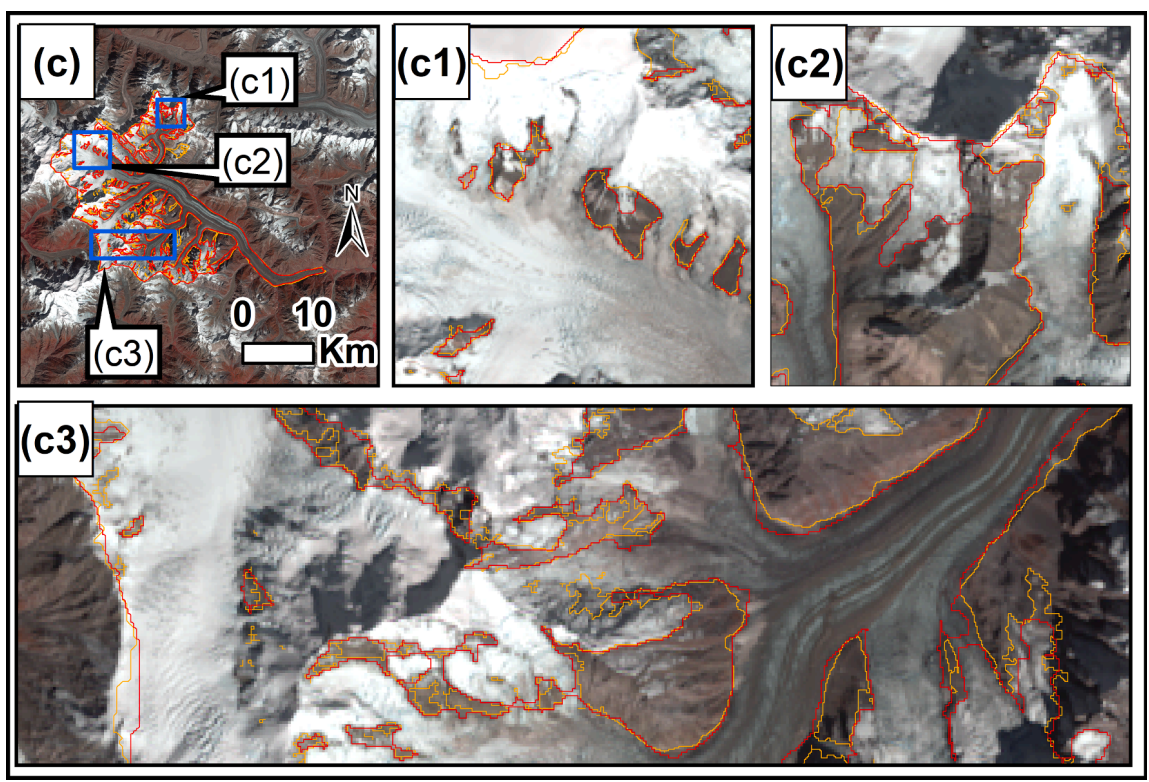

**Fig. 10c.** The boundary covered both ablation zone and SCAZ of Chogo Lungma glacier. The final output was processed by region size thresholding to remove residuals of small (i.e., few pixels), non-snow areas. The red and gold boundaries indicate the GlacierNet2 and reference, respectively.

*3.1. Ablation zone mapping*

The primary strategy for improving ablation zone mapping here was to effectively utilize the advantageous properties of different CNN models. Our results indicated that GlacierNet and DeepLabV3+ models produced similar boundary predictions, as reflected by their comparable IOU scores (Table 1). The former method produced a higher recall score and tended to overestimate the termini boundaries, whereas the latter maintained a superior precision score while underestimating termini locations. GlacierNet represents a computationally light and simplified model with a lower similarity requirement than the more complex DeepLabV3+ model. Lowering the similarity requirement to include more positive case pixels assists with incorporating the higher spatial heterogeneity and complexity of DCGs. Furthermore, this approach helps capture the positive cases with lower similarity, although this is associated with slight overestimates of boundary locations, particularly in the complex terminus regions. Alternatively, the more complex model of higher similarity requirements underestimates glacier boundaries (Fig. 11a1, c1). Moreover, both networks performed well in the ablation zone (excluding terminus), although DeepLabV3+ was slightly more accurate (Fig. 11b1, b2, d1, d2). The fused network utilized 256 and 32 feature maps from DeepLabV3+ and GlacierNet, respectively. Although the convolutional layer objectively weighted the input feature maps through training, more maps often occupy the greater predictive weight. Accordingly, DeepLabV3+ dominated the decision-making process, whereas GlacierNet played a more supplementary role. The fused network incorporated and improved the precision of identifying boundaries in the ablation zone estimates, whereas post-processing primarily increased recall scores. The network fusion strategy, for terminus location, improves some of those predictions, but also results in some underestimations (Fig. 12D-2) because of the disagreement between networks.

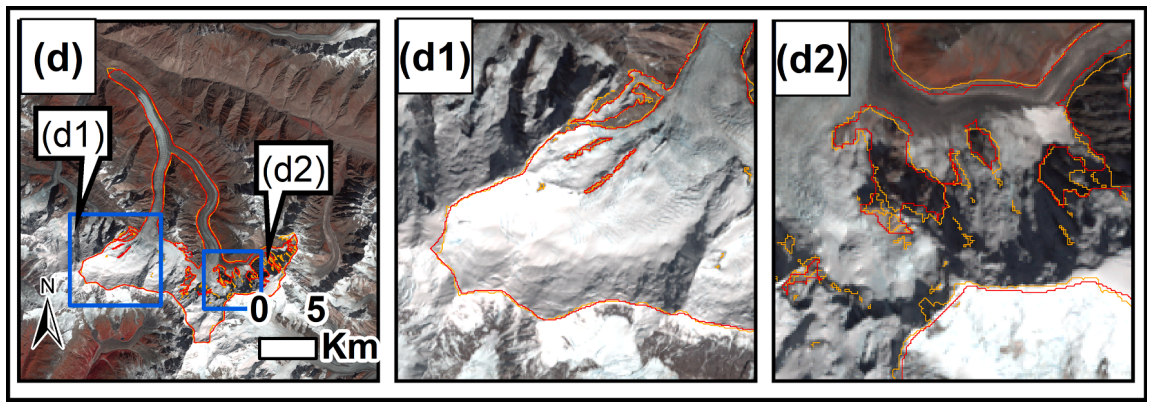

**Fig. 10d.** The boundary covered both ablation zone and SCAZ of the Barpu glacier. The final output was processed by region size thresholding to remove residuals of small (i.e., few pixels), non-snow areas. The red and gold boundaries indicate the GlacierNet2 and reference, respectively.

Additionally, this strategy contributed to the sample data definition for KNN, as the highest precision score indicated the highest accuracy of true-positive cases.

The KNN was applied to each uncertain terminus zone as determined by the IOU score of *E-1* and *E-2*, which is indicative of the positive agreement between the two. If both networks tended to over- or underestimate the terminus location with a high level of agreement, the estimation was deemed most likely appropriate or containing only minor errors (e.g., Fig. 12, termini 2 and 5; Table 2). Conversely, estimates were deemed inaccurate when low levels of agreement were observed (e.g., Fig. 12, termini 1, 3, and 4; Table 2). This uncertain terminus zone identification strategy was implemented to address the limitations of accuracy assessment without actual ground verified data.

Generally, ML algorithms require manually collected target sample data. To derive a more universally applicable approach and to classify the complex and variable glacier termini, the local and automatic sample data selection strategy (i.e., sampling every glacier, respectively) for KNN was utilized. The sample data selection strategy systematically and effectively defined positive (glacier pixels) and negative (non-glacier pixels) samples, identifying the former using the AND operation for enhanced certainty, as the fused network maintained the highest level of precision for judging positive cases. Similarly, the negative samples were selected from the pixels, which both networks identified as belonging to the negative category. Notably, the positive or negative cases were geographically adjacent to uncertain pixel clusters, thus informing through their high levels of similarity to pending pixels.

The KNN classifier uses information that represents object similarity features for classification. This is the key factor that affects classification performance. The utilized feature data in this study to describe an object were CNN-generated automatic feature maps (using supervised learning strategies) rather than original data or those extracted from image processing algorithms. On the other hand, the classical image processing algorithms generate specifically designed feature descriptors for representing an object. Comparatively, the feature data that are generated by well-trained CNN kernels cover more comprehensive and objective information than those from specifically designed feature descriptors, particularly the multispectral data. It becomes more challenging to account for each type of spectral data when designing feature descriptors. In addition, the feature data used in KNN were not processed through any down-sampling, so each pixel had a corresponding feature vector containing neighboring information from superimposing the reception fields of the three convolutional layers. Furthermore, the utilized CNN feature data are low-level features generated from the first few layers rather than the high-level features from the last several layers of CNN. Therefore, the low-level features are influenced less by CNN's decision, and are more neutral. Since the purpose of utilizing KNN is to correct the CNN's misclassification, the biased features influenced by CNN can mislead the KNN's decision.

When KNN is applied directly without assistance from CNN models' estimation, the resulting classification can be inaccurate, as the former improves terminus estimation, and the latter provides feature data and references for effectively selecting sample data across the study regions. Additionally, we used many different KNN classifiers created only using local sample data for every corresponding terminus, rather than a classifier that includes all the sample data. This helps in avoiding the influence of samples from the other glaciers. Other ML algorithms, such as support vector machines and random forests, are more time-consuming and computationally intensive due to training requirements while producing similar or slightly poorer results. Also, our analysis shows that the K-value variations cause a minor effect on the evaluation indices as KNN is only employed in some termini regions (Table 3).

Because the KNN was applied to the XOR region of two CNN estimates (*E-1* and *E-2*), GlacierNet2 was limited by the false-negatives identified by GlacierNet and DeepLabV3+ networks. Possible solutions can be explored by modifying network structure to help enhance true-positive cases, as well as acceptable false-positives. Like the GlacierNet design optimization, the modification requires a systematic experiment to tune the network structure parameters. For example, CNN models are generally designed for RGB images taken from different angles, poses, and depth, but the satellite imagery has an almost fixed angle, depth, and pose. Therefore, it does not require the 3-dimensional object and depth understanding of CNN models (Xie et al., 2021). Also, multispectral satellite data offers more surface characteristics at different wavelengths to differentiate the target object, than the RGB images. Consequently, the CNN structure optimization experiment aimed at processing satellite imagery and mapping glaciers can effectively improve the CNN models. Furthermore, the incorporation of additional CNN models can potentially construct more comprehensive unions, thus revealing more true-positive cases. Network estimates with different or even identical structures will maintain different levels of accuracy even after training with the same data, thus being capable of complementing each other. Even if the network is modified or more network structures are utilized, some pixels may not be sufficiently similar to the training data, which themselves may be misclassified; therefore, in this case, the selection of additional training data (primarily focusing on the terminus region) is a feasible solution.

Furthermore, comparing the GlacierNet2 produced outlines with reference outlines in the training region shows that GlacierNet2

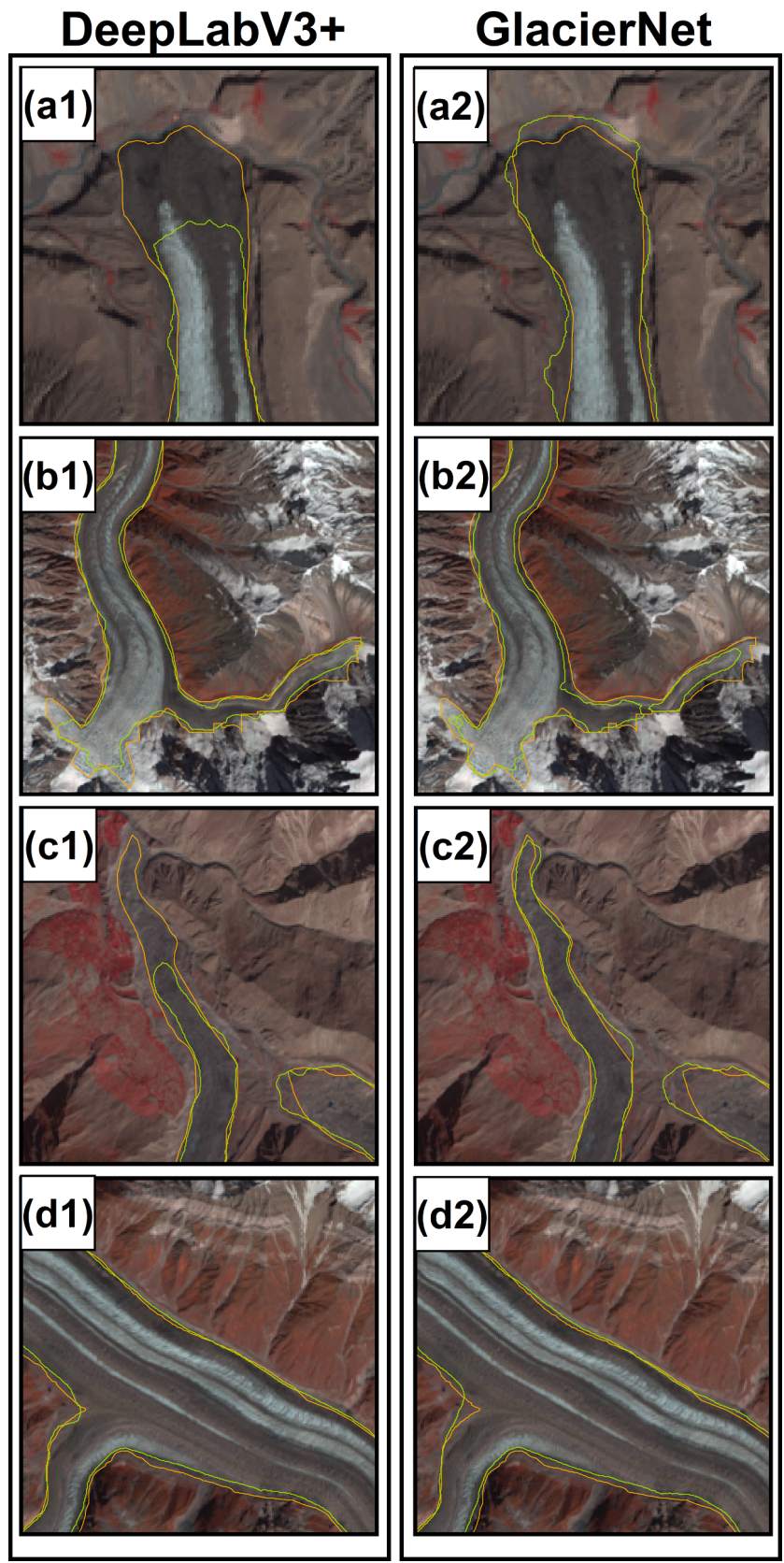

**Fig. 11.** *Comparing DeepLabV3+(a1-d1) and GlacierNet (a2-d2) generated boundaries. Green and gold color boundaries indicate the networks generated boundaries and references, respectively* (Xie et al., 2021).

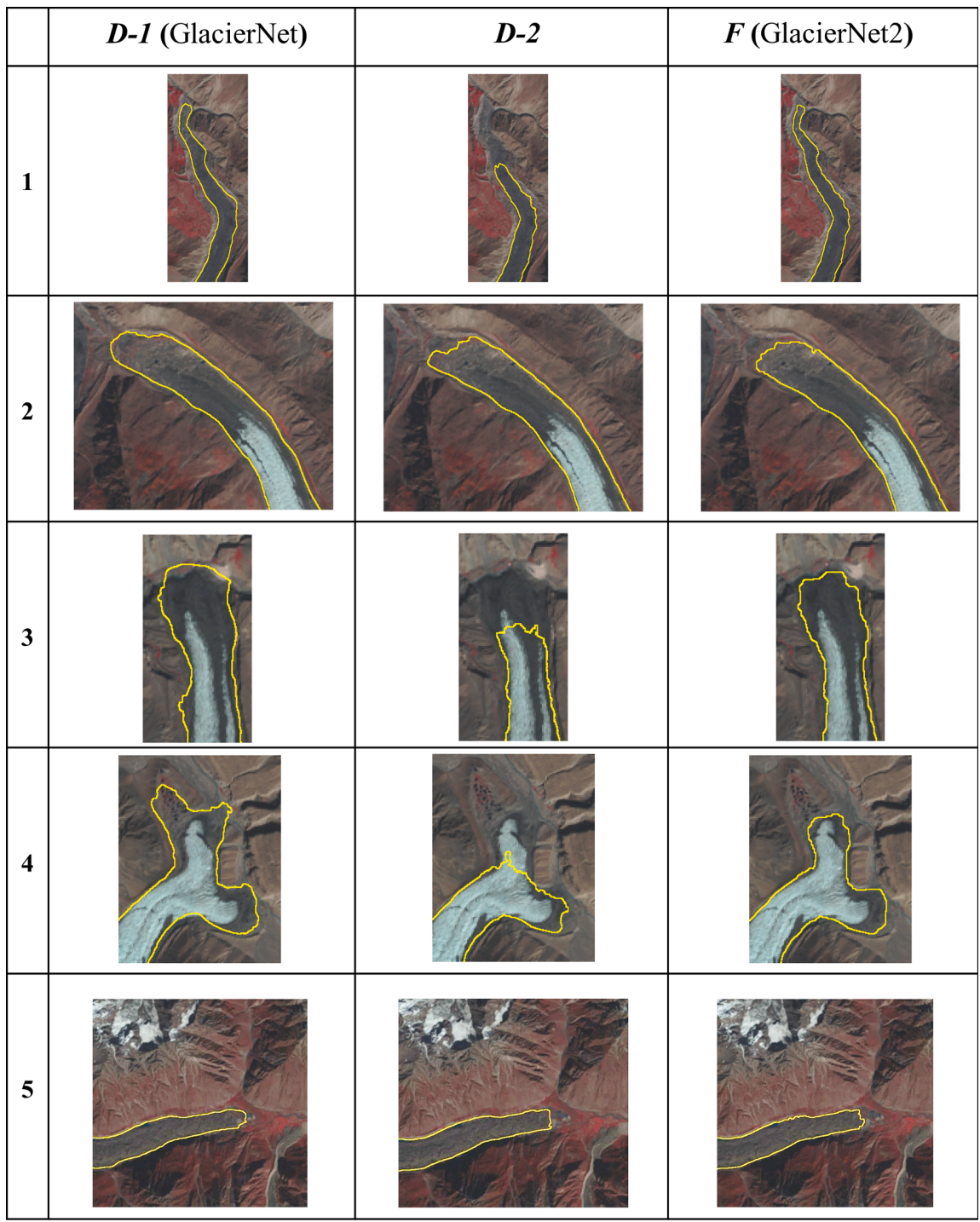

**Fig. 12.** Terminus samples for comparing the *D-1* (GlacierNet), *D-2*, and *F* (GlacierNet2). Estimates of termini 1, 3, and 4 maintained larger levels of disagreement in *D-1* and *D-2*; thus, they were improved upon by both KNN processing and vegetation zone removing. Estimates of termini 2 and 5 held high levels of agreement between *D-1* and *D-2*, requiring only the removal of the vegetation zone from *D-2*.

**Table 2**
IOU comparison of terminus samples in Fig. 12.

|   | IOU (*D-1, D-2*) | IOU (*D-2, F*) | IOU (*D-1, F*) |
|---|---|---|---|
| 1 | 0.6791 | 0.7704 | 0.8779 |
| 2 | 0.9171 | 0.9683 | 0.8900 |
| 3 | 0.4952 | 0.6004 | 0.8221 |
| 4 | 0.5840 | 0.7561 | 0.7782 |
| 5 | 0.9148 | 0.9426 | 0.9619 |

**Table 3**
KNN's K-value evaluation indices for debris-covered glacier ablation zone mapping.

| K-value | IOU | RC | PC | SP | FM | ACC |
|---|---|---|---|---|---|---|
| 1 | 0.8838 | 0.9009 | **0.9791** | 0.9944 | 0.9383 | **0.9735** |
| 5 | **0.8839** | **0.9010** | 0.9790 | 0.9944 | **0.9384** | 0.9735 |
| 10 | 0.8838 | 0.9008 | 0.9791 | 0.9944 | 0.9383 | 0.9734 |
| 20 | 0.8838 | 0.9009 | 0.9790 | 0.9944 | 0.9383 | 0.9734 |

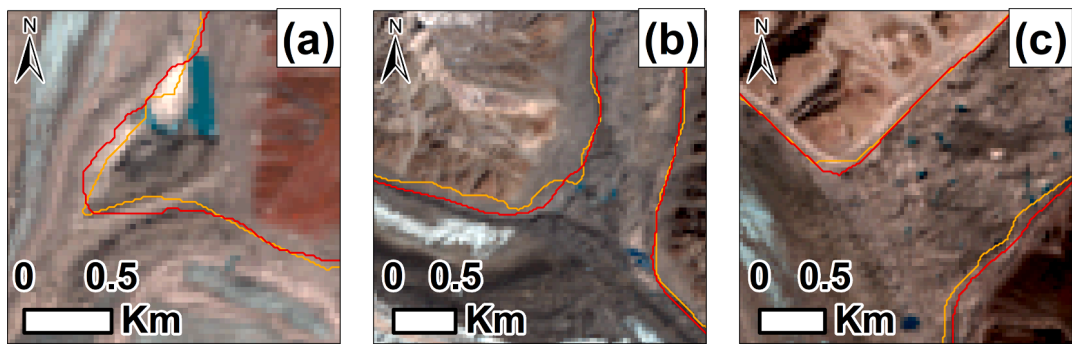

**Fig. 13.** Comparison of GlacierNet2 (red) and training data reference (gold).

produces higher quality boundaries even after learning the training data that contains some error, as discussed in section 2.1 (Fig. 13).

Overall, the ablation zone improvements are highly dependent on the multi-model association, which helps 1) keep the dominant capabilities of DeepLabV3+ in delineating the ablation zone (excluding terminus) boundaries, 2) identify the low similarity glacier pixels, especially near the terminus, via introducing the GlacierNet with a lower similarity requirement to input data, 3) cross-references (i.e., terminus IOU comparison between CNNs) discover the potentially inaccurate terminus estimations, 4) define the positive and negative samples of each unique KNN classifier used for reconsidering the pixel category of the potentially inaccurate terminus, 5) designate and limit the reconsideration area (pending area) for the KNN classifier, and 6) improve the proportion of true-positive in positive cases to assist the sample data definition for KNN.

*3.2. Snow-covered accumulation zone mapping*

GlacierNet2 classified the SCAZ by assembling DB estimates from multiple aspects. The intermediate output *G-1* underestimated a few glaciers in the study area (e.g., Virjerab glacier, Fig. 14a), whereas intermediate *G-3* overestimated due to the inclusion of indirectly connected snow regions that also flow into the ablation zone (Fig. 14b). Consequently, when both *G-1* and *G-3* were incorporated, the most accurate boundary estimates of the Virjerab glacier were derived (Fig. 14c).

The accuracy of SCAZ estimates is highly associated with the quality of ablation zone mapping results. The basic principle employed was to search the flow path starting from the ablation zone in order to identify the snow-covered area. The DB segment, however, may have been undetected due to incomplete ablation zone. The inherent heterogeneity of glacier systems and complexity of surface topography, with protruding obstacles, such as ice cliffs or seracs, sometimes inhibits accurate DB estimation. Removing them through thresholding can mislead DB estimates. To minimize the effects from protruding obstacles, regional averaging by elevation is necessary, although it does not thoroughly address DB loss. Therefore, the snowline was adjusted to the higher elevation value compared to the two network estimates. In addition, the SCAZ mapping performance can be improved with high-resolution DEM and ablation zone estimations, as incorrect altitudes could cause the derived algorithm to under- or over-allocate basin segments.

Moreover, the GLIMS boundary was generated using satellite data acquired at a different time than the satellite data employed here, and the variability of the SCAZs over time may mean that GlacierNet2 performance may have been underestimated. Variations in sensor data spatial resolution, preprocessing, and registration errors can also determine predicted locations.

*3.3. Data issues and improvement*

Our results indicate that DCG mapping issues that have been reported in the literature can potentially be addressed using neural computing and CNN models (Bishop et al., 2007; Bolch et al., 2007; Bolch and Kamp, 2005; Mölg et al., 2018; Paul et al., 2004; Racoviteanu et al., 2009; Rastner et al., 2013; Robson et al., 2016; Shukla et al., 2010b, 2010a; Xie et al., 2020). Glacier boundaries for many glaciers in central Karakoram were found to be comparable in terms of their general location compared to reference data, which has been primarily obtained by human interpretation, empirical statistical analysis, and in many cases, post-processing of classification results. However, it is important to note that the reference data utilized does contain errors and uncertainties when compared to results from experts who have direct field experience and data collected from the field. This is because traditional approaches to glacier mapping cannot account for the complex geometry and topographic property variation caused by climate-glacier dynamics that occur within this region. Furthermore, sensor system characteristics, radiometric calibration, orthorectification, preprocessing, and terrain parameter extraction algorithms represent other potential issues that need to be accounted for when comparing results to reference data.

In addition to the original data/information input that feeds into the model for training and prediction, one of the key issues in evaluating the use of neural computing involves understanding the nature of the information that various CNN models are using. This has historically been labeled the "Black Box problem" as information extracted from the data and structure of the model generates feature input that can be used for decision making, as is the case in our work. The question becomes: what is the nature of the information that is required and used to produce meaningful estimates of boundary locations. When spectral and terrain information are utilized, what form of information dominates, and more specifically, what is the nature of that information. Is it spatial geometry information and is the information scale-dependent? Is it possible that spatial topological information such as landcover and terrain spatial structure are being utilized, or does the architecture generate other emergent properties of the landscape that exhibit their own scale dependencies that have not been formally quantified or conceived of yet. This represents a significant problem with the use of deep learning and more complicated ANN architectures, and more research involving the use and characterization of spatial information generated from neural computing for improved understanding and mapping of complex landforms is sorely needed.

It is also important to note that more research is required to ascertain the essential types of data and information that can be used in neural computing efforts to optimize and improve upon mapping DCG. From a spectral perspective, there is a need to evaluate the utility of radiometric calibration, and accounting for coupled atmospheric-topographic effects that govern the spectral variability in satellite imagery. Although ratioing is commonly utilized to reduce topographic effects, they are not completely removed and numerous researchers have indicated that multi-scale effects need to be accounted for in mapping efforts (Bishop et al., 2019). It would also be important to remove multicollinearity in the spectral input data.

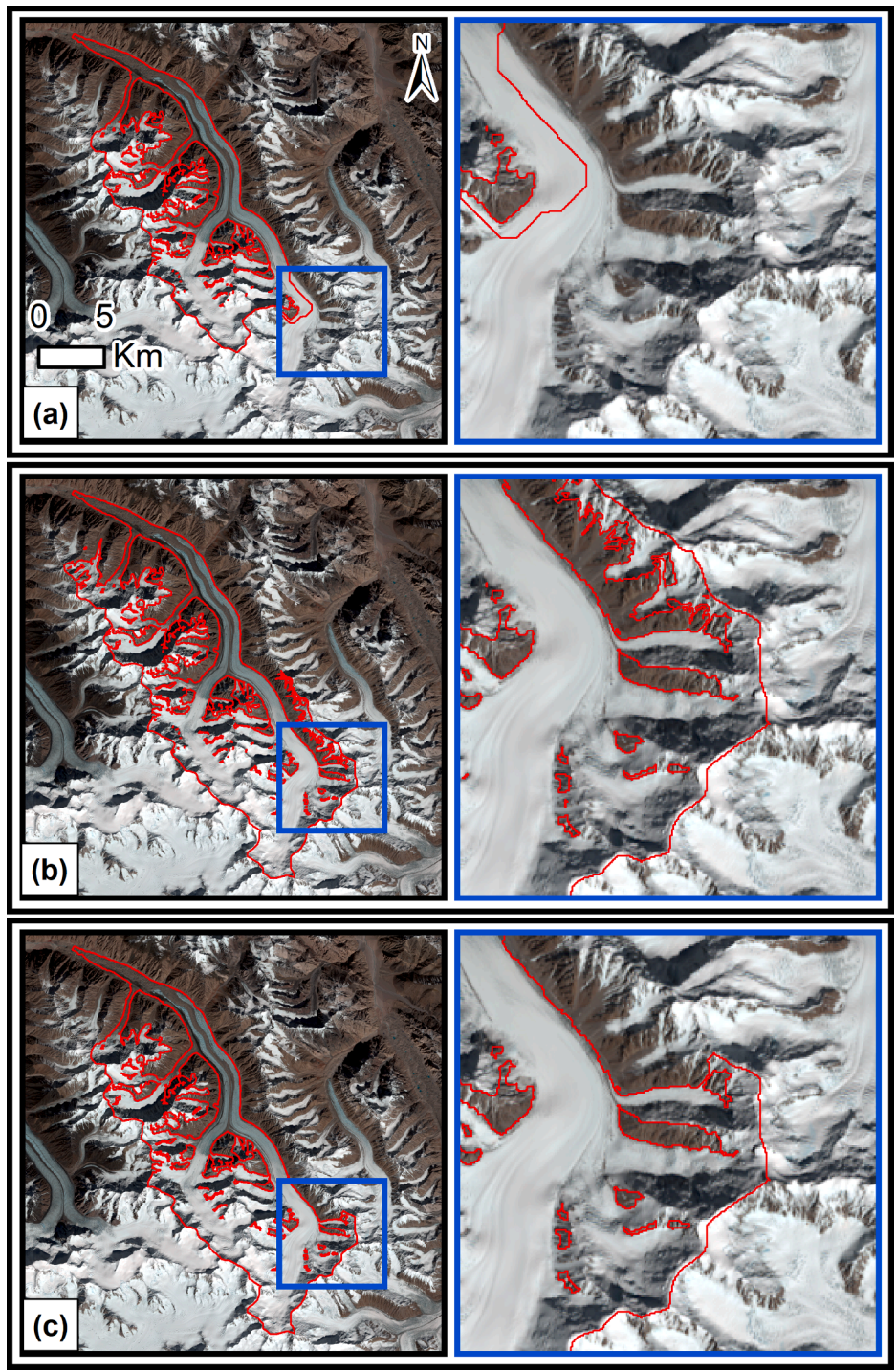

**Fig. 14.** Virjerab glacier, an illustrative example for comparing the final and intermediate outputs: (a) Intermediate output *G-1*; (b) Intermediate output *G-3*; and (c) The final output processed by region size thresholding to remove small (few pixels), non-snow areas.

From a terrain analysis perspective, it is important to note that the usage of the selected LSP may not be optimal for characterizing glacier boundaries, as many different algorithms exist to produce an estimate of surface topographic properties. Furthermore, there are many additional properties that are scale-dependent that may greatly improve boundary identification. Similarly, terrain structure information related to hillslopes, valley floors and eminences can also be used to spatially constrain and identify glacier surface features that can promote more accurate delineation of glacier landform elements and boundaries. This higher-level terrain information, usually generated from object-oriented analysis, would greatly improve mapping results because the spatial organizational structure of the landscape and glaciers have unique patterns that should be identified using a CNN. In this way, further refinement and an improvement in our results is possible, thereby decreasing the uncertainty associated with boundary locations, regardless of the structural complexity of the landscape.

## 4. Conclusion

An updated version of the GlacierNet approach, GlacierNet2, presented here uses Landsat images, DEM, LSP, and incorporates multi-model learning and basin level hydrological flow information to automatically map debris-covered glaciers, including the ablation and accumulation zones. GlacierNet2 improves the ablation zone mapping with a higher IOU score of 0.8839. Of which the advanced network structure contributed to the improvement of mapping ablation zone lateral boundaries, and the automated post-processing helped in improving terminus estimation. Also, the GlacierNet2 incorporates open-source TopoToolbox functions to include an accumulation zone for a complete glacier mapping. The overall IOU score for the classification of the entire glacier basin is 0.8619, which reflects the strong ability of GlacierNet2 to use limited training data to achieve high-performance estimates of glacier boundaries. Specifically, it reaches high accuracy by accounting for the (1) different similarity requirements and structures of CNN models and their ability to class ablation zone and extract data feature; (2) the effectiveness of KNN to the low amount of data input, the fast and non-parametric properties of KNN, and the feature classification ability of KNN; (3) the high processing speed, user-friendly platform, flexible properties, and the improved function of drainage basin estimation of TopoToolbox. Although the GlacierNet2 can accurately delineate boundaries and offers a tool for researchers to automatically map glaciers, there is room for improvement by addressing data quality, model structure, and post-processing algorithm issues. The proposed approach can also be improved by utilizing multi-temporal data and spatially extending the training to include a wide variety of glacier morphological characteristics.

*CRediT authorship contribution statement*

**Zhiyuan Xie:** Conceptualization, Methodology, Software, Writing – original draft, Writing – review & editing. **Umesh K. Haritashya:** Conceptualization, Methodology, Writing – original draft, Writing – review & editing. **Vijayan K. Asari:** Methodology, Writing – original draft, Writing – review & editing. **Michael P. Bishop:** Writing – original draft, Writing – review & editing. **Jeffrey S. Kargel:** Writing – original draft, Writing – review & editing. **Theus H. Aspiras:** Writing – original draft, Writing – review & editing.

**Declaration of Competing Interest**

The authors declare that they have no known competing financial interests or personal relationships that could have appeared to influence the work reported in this paper.

**Acknowledgement**

ZX thanks funding support from the Integrative Science and Engineering Center, College of Arts and Science, and School of Engineering at the University of Dayton. UKH and JSK was supported by the NASA High Mountain Asia grant 80NSSC19K0653 and NASA Interdisciplinary Research in Earth Science grant 80NSSC18K0432. UKH was also supported by the University of Dayton's Mann Endowed Chair in the natural sciences. We also thank the Ohio Supercomputer Center (www.osc.edu) for extending allocation hours through project PNS0435 to run our GlacierNet algorithm. Landsat OLI images are acquired from the U.S. Geological Survey (https://glovis.usgs.gov) and the ALOS Global Digital Surface Model (AW3D30) by the Japan Aerospace Exploration Agency. We also thank the original authors and developers of TopoToolbox and Brennan W. Young for developing original geomorphometric layers for GlacierNet.

**Appendix A. Supplementary material**

Supplementary data to this article can be found online at https://doi.org/10.1016/j.jag.2022.102921.